\documentclass{iopart} 

\usepackage{array}
\usepackage{graphicx}

\begin{document}

\title[Phase and alignment noise in grating interferometers]{Phase and alignment noise in grating interferometers}

\author{A Freise}
\address{School of Physics and Astronomy, University of Birmingham, Edgbaston, Birmingham B15 2TT, UK}
\ead{adf@star.sr.bham.ac.uk}
\author{A Bunkowski}
\address{Institut f\"ur Gravitationsphysik, Leibniz Universit\"at Hannover and Max-Planck-Institut 
f\"ur Gravitationsphysik (Albert-Einstein-Institut), Callinstr.~38, 
30167 Hannover, Germany}
\author{R Schnabel}
\address{Institut f\"ur Gravitationsphysik, Leibniz Universit\"at Hannover and Max-Planck-Institut 
f\"ur Gravitationsphysik (Albert-Einstein-Institut), Callinstr.~38, 
30167 Hannover, Germany}

\begin{abstract}
Diffraction gratings have been proposed as core optical elements
in future laser-interferometric gravitational-wave detectors.
In this paper we derive equations for the coupling between alignment noise and phase 
noise at diffraction gratings. 
In comparison to a standard reflective component
(mirror or beam splitter) the diffractive nature of the gratings causes an additional 
coupling of geometry changes into alignment and phase noise.
Expressions for the change
in angle and optical path length of each outgoing beam are provided as 
functions of a translation or rotation of the incoming beam with respect
to the grating. The analysis is based entirely on the grating equation and the geometry of the
setup. We further analyse exemplary optical setups which have been proposed for the use in future
gravitational wave detectors. We find that the use of diffraction gratings
yields a strong coupling of alignment noise into phase noise.
By comparing the results with the specifications of current detectors we show that this
additional noise coupling 
results in new, challenging requirements for the suspension and isolation systems for the optical
components. 
\end{abstract}



\maketitle

\setlength{\extrarowheight}{0.2cm}

\section{Introduction}

The search for gravitational waves has led to a new class of extremely
sensitive laser interferometers.  The first generation of large-scale 
laser-interferometric gravitational wave detectors \cite{geo2006, virgo2006short, ligo2004, tama2004} 
is now in operation with the aim of accomplishing the first direct detection 
of gravitational waves. Simultaneously, new interferometer concepts are
evaluated for future detectors. 

Traditionally, partly transmissive mirrors are used in interferometers to split
and
combine coherent optical light fields.
For high precision laser interferometers, such as for gravitational wave
detection, non-transmissive reflection gratings offer a useful alternative way
of splitting and combining.
The resulting all-reflective interferometers are beneficial because, firstly,
they reduce the impact of \emph{all} thermal issues that are associated with absorbed laser
power in optical substrates and, secondly, they allow for opaque materials with
favourable mechanical and thermal properties.
With these two qualities all-reflective interferometer concepts have, in
principle, great potential to become key technologies for enhancing the
sensitivity of future generations of laser interferometric-gravitational wave
detectors.
From a functional viewpoint every partly transmissive mirror within an interferometer
can be substituted by an appropriate reflection grating because of its analog
input-output phase relations.
However, the geometry of the interferometer changes considerably when
diffraction gratings are used.
In this context several interferometer concepts based on gratings have been
proposed~\cite{Drever96,Schnabel} and some of them have been demonstrated
experimentally \cite{SB98,BBBDS04,BBCKTDS06}. Also the influence of a grating
structure on the mechanical quality factor of a test mass has been studied
\cite{Nawrodt07}.
Here we investigate how certain peculiarities of grating interferometers affect their
ability to reach high strain sensitivities.
In particular we derive formulae for the alignment noise of such
interferometers.
Grating movements within these interferometers or beam movements on the grating
affect the phase of the light differently than movements of mirrors and 
beam splitters in conventional interferometers.
This is due to the reduced symmetry that diffraction gratings show compared to
 mirrors.
Usually, the test masses in gravitational wave detectors show cylindrical
symmetry, therefore their roll movement is of no concern.
Gratings are merely invariant against translational displacement in direction
parallel to the grating grooves, but certainly not for rotation.
Therefore roll movement can be considered an additional degree of freedom that
will be treated here.
Moreover, a translational displacement of a grating parallel to its surface in
direction perpendicular to the grating grooves will induce a phase
shift~\cite{Wise05} to the reflected light.
%

After a brief review of known all-reflective interferometer concepts we derive
analytic expressions that describe the phase effects for various motions of
optical components in the particular interferometer and compare them to the
well known ones for conventional interferometers.

\subsection{Gratings as functional optical elements in interferometers}\label{sec:gratings}
\begin{figure}[htb]
    \centerline{\includegraphics[width=5.cm]{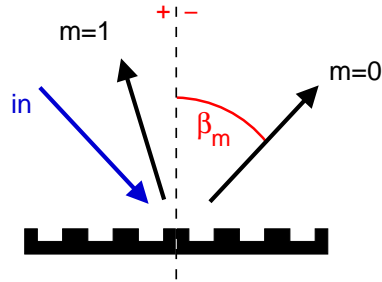}}
    \caption{A grating illuminated by a beam (in). The number of outgoing beams
    is given by the grating equation~(\ref{eq:gr}). The beams are numbered by an integer
    (m) and the angles
     with respect to the grating normal are given as $\beta_m$.
    The angle of the incident light is $\alpha=-\beta_0$. Shown is a non-Littrow mount.}
    \label{fig:gr}
\end{figure}

A surface with a  periodic modulation  of optical properties, so-called
grooves, defines a diffraction grating.
Let's have a look at~Figure~\ref{fig:gr} and  consider incident light of
wavelength $\lambda$ in the plane perpendicular to the grating grooves and its
surface.
For a grating period $d$ and  an incidence angle of $\alpha$, measured from the
grating normal, the angle $\beta_m$ of the $m$th diffraction order is given by
the well-known grating equation
%
\begin{equation} \label{eq:gr}
    \sin\alpha + \sin\beta_m = m\lambda/d.
\end{equation}
For transparent materials the orders will exist in transmission and reflection.
One obtains an all-reflective beam splitter when the grating is combined with a
high reflectivity coating, or transmitted orders are suppressed by some other
means.
%
The existence of higher orders depends on the choice of $d$ and $\alpha$.
For our purposes only one or two additional orders are required, so that
$d\sim\lambda$.

\begin{figure}[htb]
\begin{minipage}{0.55\textwidth}
    \centerline{\includegraphics[viewport=0 -150 647 345,scale=0.2]{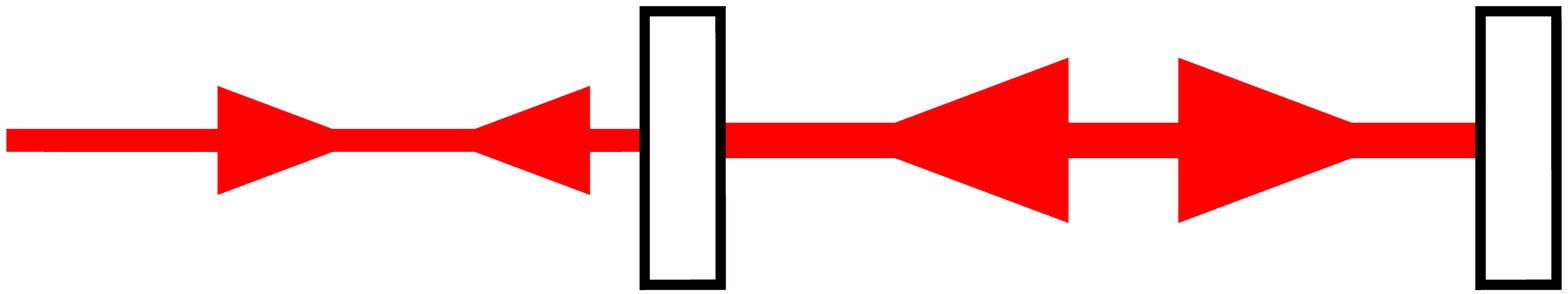}}

\vspace{2mm}
    \centerline{Linear Fabry Perot Cavity}

\vspace{5mm}
    \centerline{\includegraphics[scale=0.2]{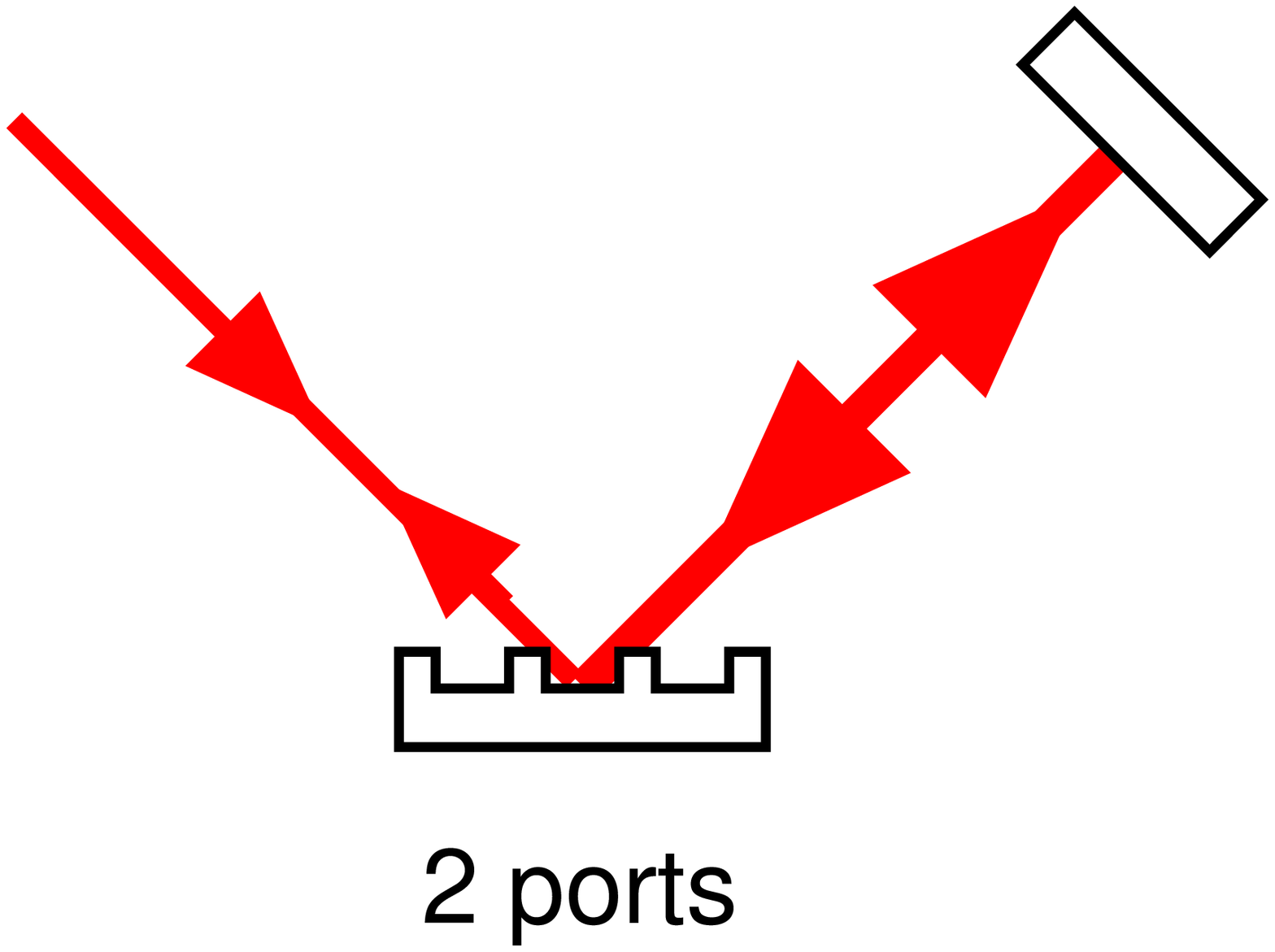} \hspace{6mm}\includegraphics[scale=0.2]{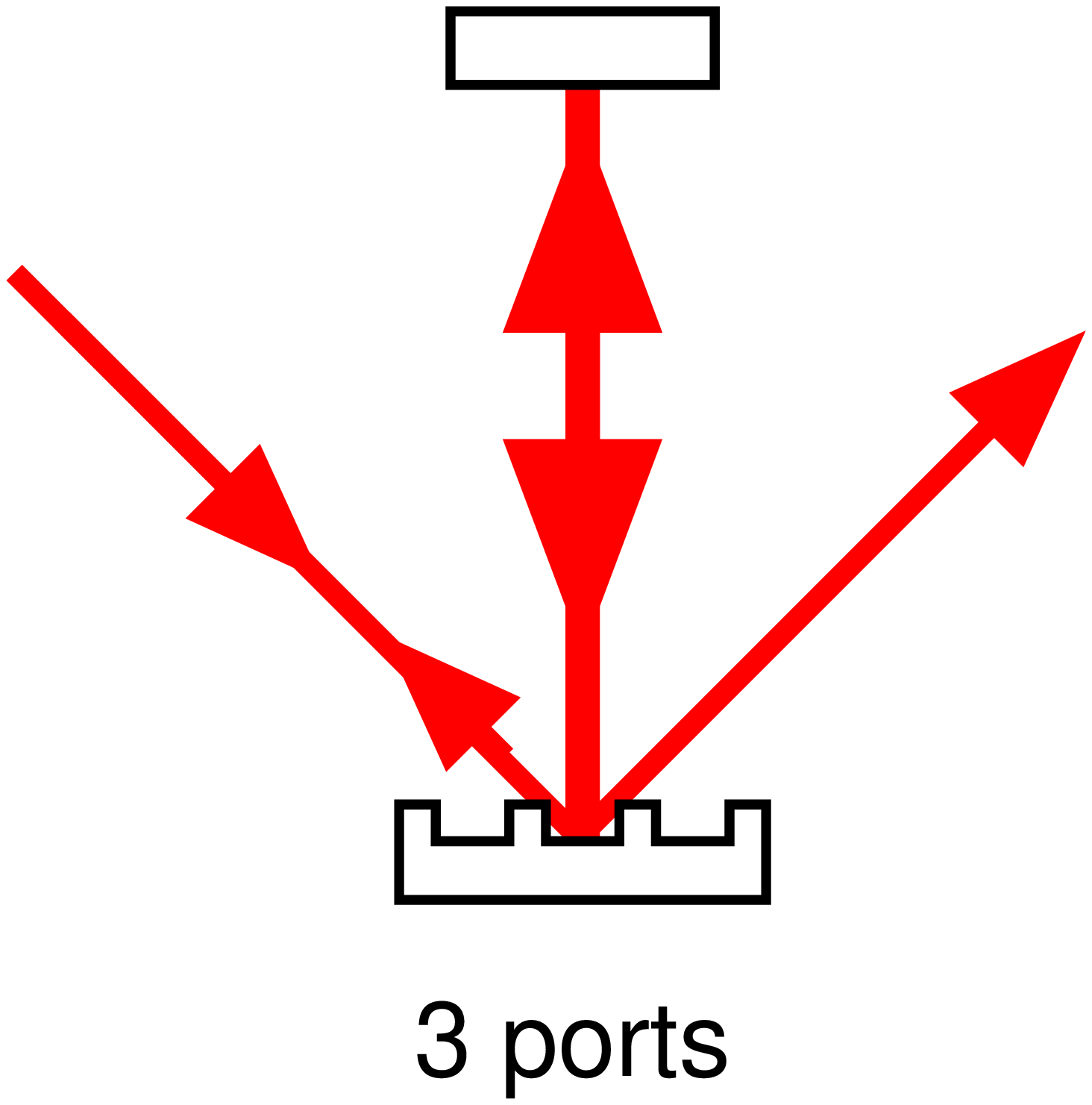}}
\end{minipage}
\hfill
\begin{minipage}{0.35\textwidth}
    \centerline{\includegraphics[viewport=0 0 647 495,scale=0.2]{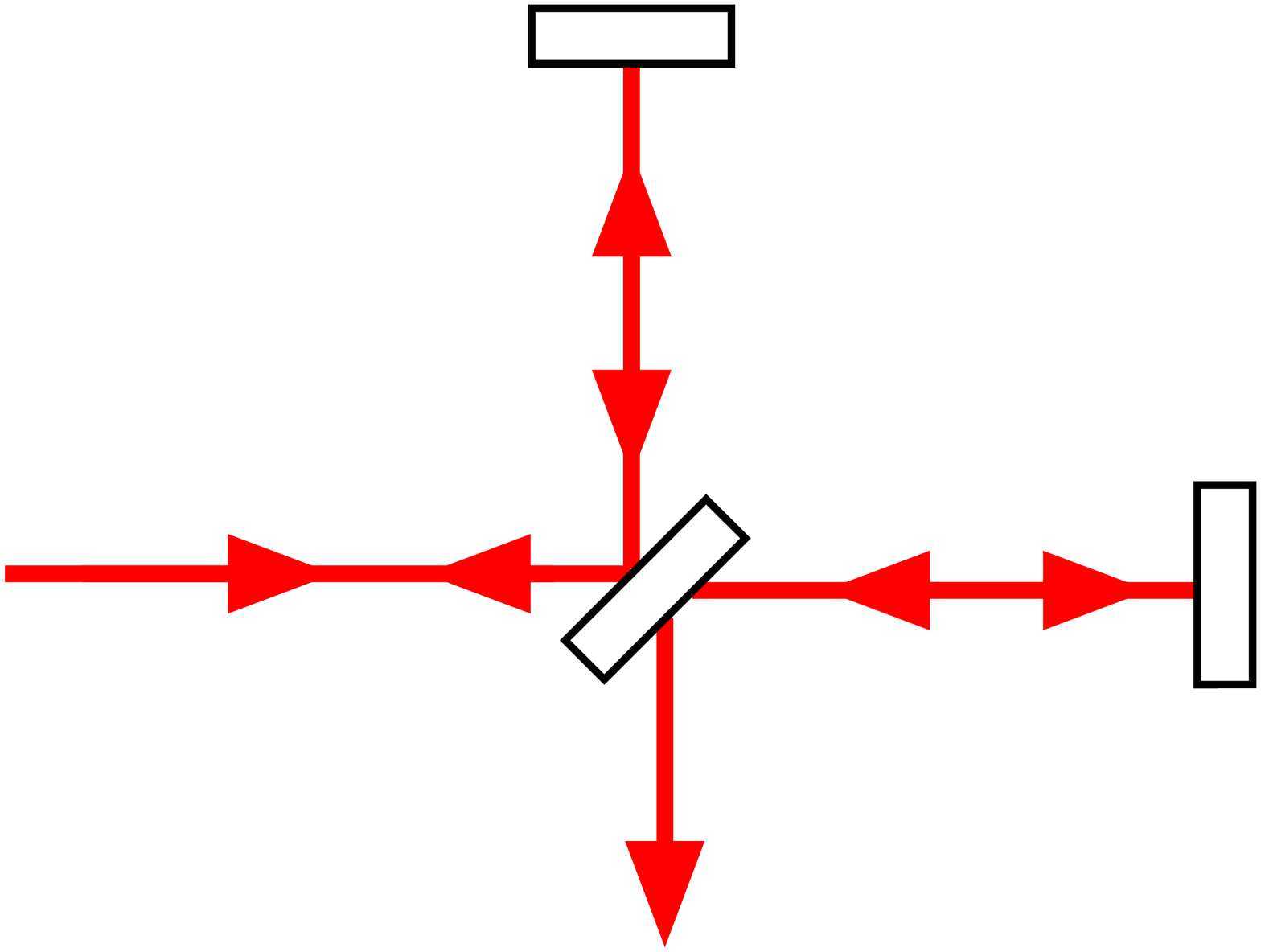}}

\vspace{2mm}
    \centerline{Michelson Interferometer}

\vspace{5mm}
    \centerline{\includegraphics[scale=0.2]{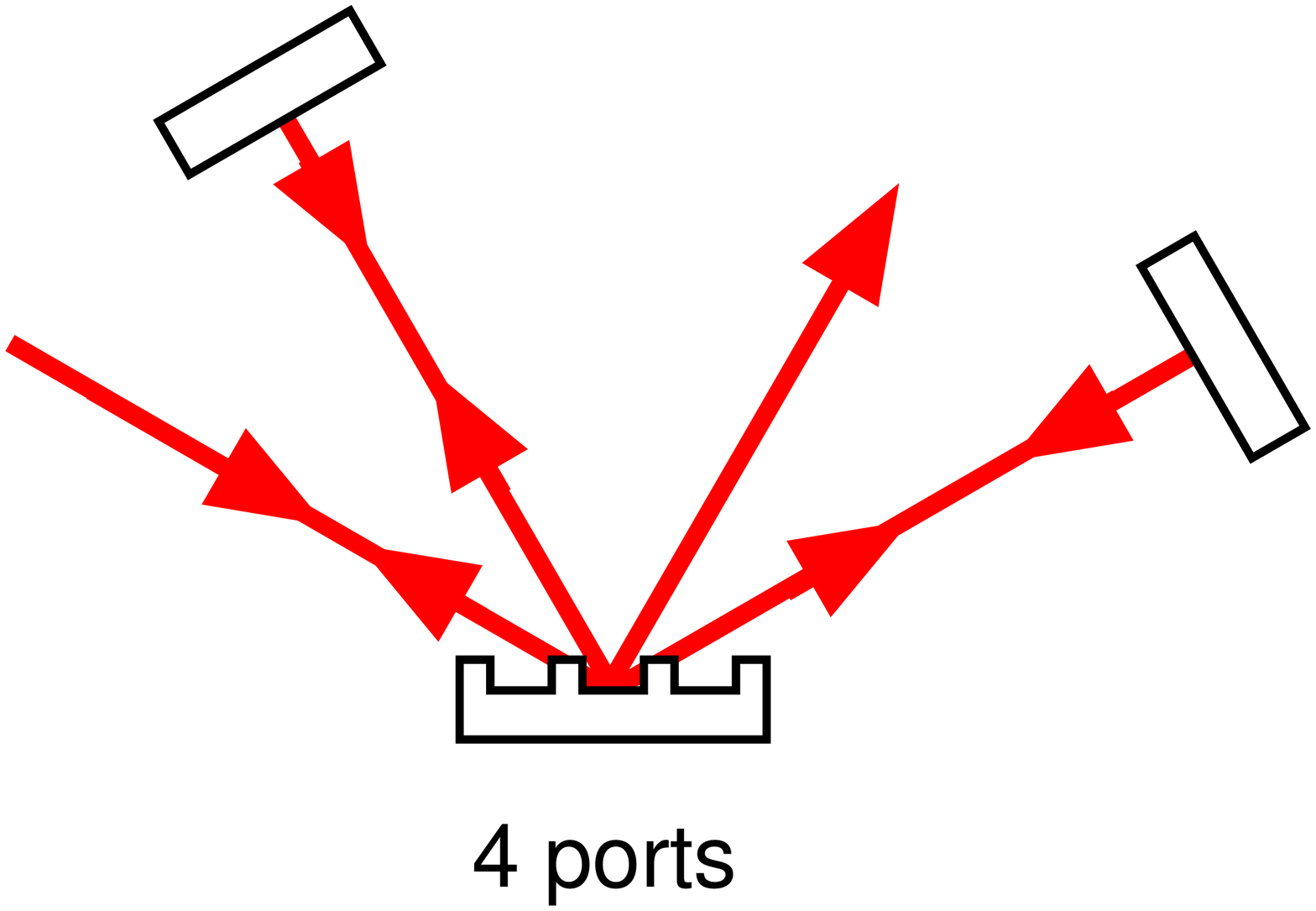}}
\end{minipage}
    \caption{(top) Sketch of a Michelson and a linear Fabry-Perot interferometer with transmissive optical
    elements and
    (bottom) possible all-reflective realisations of these devices based on diffraction gratings. Note that the Fabry-Perot
    interferometer can either be realised with a grating in first-order (resulting in two ports) or second-order Littrow
    mount (three ports).} \label{fig:ifos}
\end{figure}

For appropriately chosen parameters there is only one additional diffraction
order and no degeneracy of ports $(\alpha\neq\beta_1),$ thus one obtains the
analog to a four-port mirror. 
This device enables, for instance, an all-reflective version of a Michelson
interferometer
as shown in Figure~\ref{fig:ifos}, provided that the efficiencies for the
specular reflection and for the diffraction into the first order are roughly
the same.

The analog to a transmissive mirror with two ports (in the case of normal
incidence) is given by a first-order Littrow configuration.
In this case also only one additional order exists
but the diffracted beam coincides with the incoming beam $(\alpha=\beta_1)$.
An all-reflective linear Fabry-Perot interferometer can be
constructed, also shown in Figure~\ref{fig:ifos}.
The maximal finesse of such a cavity is limited by the first-order diffraction
efficiency of the grating that is used to couple light to the cavity.

Parameters can likewise be chosen to allow for a second-order Littrow
configuration (two additional orders and $\alpha=\beta_2$). This 
results in a beam splitter with three ports, which can also be used to 
construct a linear Fabry-Perot interferometer
(Figure~\ref{fig:ifos}).
Its maximal finesse is limited by the specular reflectivity of the grating
rather than the diffraction efficiency.
Such a three-port splitter has no simple analog to a conventional transmissive
mirror and its input-output phase relations are more complex \cite{Bunki05}.
However, the resulting properties of such resonators are well understood and
controllable~\cite{Bunki06}.

%

\section{The geometry of the optical setup}

In this section we consider the effects of geometry changes,
namely translations and rotations of the grating or the incident
beam away from the initial, correctly aligned setup. We derive the
mathematical relation between each geometry change and the change in
optical path length, tilt and translation of the
outgoing beam. In the following we often use the word beam when referring
to the light fields interacting at the grating. However, it
should be noted that all computations are based on the
assumption of plane waves and infinite sized gratings.
Furthermore we only consider an idealised reflection grating while
any influence from a diffractive coating is neglected. The purpose of
this is to summarise the differences between the ideal grating and
ordinary (ideal) mirrors or beam splitters.

\begin{figure}[h]
    \centerline{\includegraphics[scale=0.3]{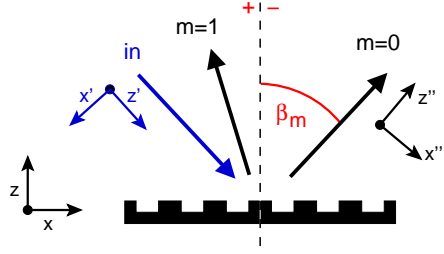}}
    \caption{Coordinate systems and angle convention at the grating} \label{fig:coordinates}
\end{figure}

Unless otherwise noted, the grating is located
in a three-dimensional coordinate system such that the impinging
beam hits the grating at the origin with the grating lying
in the $x$-$y$ plane and the grating structure (grooves) being parallel
to the $y$-axis. Thus, the nominal use of the grating requires
the incoming beam to be in the $x$-$z$ plane.
The incoming and outgoing beams can be defined by unit vectors
in the direction of propagation $\vec{p}$ and $\vec{q}$
respectively. In addition, it is useful to define coordinate systems based on the
incoming and outgoing beam in the perfectly aligned systems:
The coordinate system of the incoming beam (denoted as $x', y', z'$)
is rotated with respect to the coordinate system of the grating by an
angle $\alpha$ around the $y$-axis with $\alpha$ being the angle of incidence.
The coordinate systems of the outgoing beam will be denoted as
$x'', y'', z''$ (see Figure~\ref{fig:coordinates}). In a well-aligned system the coordinate system of an outgoing
beam is rotated with respect to that of the grating around the $y$-axis
by the angle $\beta_m$.

Any change of the geometry of the optical setup can change the angle of the outgoing beams 
as well as the longitudinal phase (optical path length). The change in optical path length 
between the initial setup and
the respective new geometry will be denoted as $\zeta$.

For a description of diffracted beams in three dimensions the
commonly used scalar grating equation is not adequate. Instead we
will base the following on the grating equation in vector
form which, for a reflective grating (in vacuum) reads:
\begin{equation}\label{eq:vec_grating}
\vec{q}\times\vec{N}-\vec{p}\times\vec{N}=\frac{m \lambda}{d}\vec{G}
\end{equation}
with $\vec{N}$ the normal vector of the grating and $\vec{G}$ the unit vector
in the direction of the grooves. In the aligned setup we can set
$\vec{N}=\vec{e}_z$ and $\vec{G}=\vec{e}_y$:
\begin{equation}\label{eq:vec_grating_aligned}
\vec{q}\times\vec{e}_z-\vec{p}\times\vec{e}_z=\frac{m \lambda}{d}\vec{e}_y
\end{equation}
We can write this in separate equations for the vector components:
\begin{equation}\label{eq:vec_grating_comp}
\begin{array}{l}
p_x-q_x=\frac{m\lambda}{d}\\
p_y=q_y
\end{array}
\end{equation}
Here, $q_z$ is not directly defined through the cross product but is given by the
definition of $\vec{p}$ and $\vec{q}$ as unit vectors.

\section{Alignment of the outgoing beam}

\subsection{Translation}\label{sec:trans}

The translations of the grating along $x$ and $y$  have no particular effect
on the geometry of the outgoing beam, only a
translation of the grating along the $z$-axis by an amount $\Delta z$ will translate the
outgoing beam along the $x''$ axis. From three triangular equations:
\begin{equation}\label{eq:trigo1}
\tan\alpha=\frac{x_1}{\Delta z}, \qquad \tan\beta_m=\frac{-x_2}{\Delta z}, \qquad \cos\beta_m=\frac{\Delta x''}{x_1+x_2}
\end{equation}
We obtain:
\begin{equation}
\Delta x''=\Delta z \cos\beta_m\left(\tan\alpha-\tan\beta_m\right)
\end{equation}

\subsection{Rotation}
There are three independent degrees of freedom for rotating the incoming
beam.
However, in this work we restrict the analysis to rotations around $x'$ and
$y'$ and neglect the influence of changes in the direction of polarization.
At the same time we want to know the effects of a rotation of the 
grating around the three axes of its coordinate system.
In the following we will first consider a rotation of the incoming
beam by $\Delta \alpha$ around the $y$ axis and $\delta'$ around
the $x'$ axis and compute the resulting rotation of the outgoing
beam ($\Delta \beta$ around $y$ axis and $\delta''$ around $x''$ axis).
Later we will derive all other alignment relations from this result.

We project the unit vector of the incoming beam 
$\vec{p}=\vec{k}/k$ to the coordinate system of the grating and get:
\begin{equation}
\begin{array}{ll}
p_x=\quad \sin\left(\alpha +\Delta \alpha\right) \cos{\delta'}\\
p_y=\quad \sin{\delta'}\\
p_z=-\cos\left(\alpha +\Delta \alpha\right) \cos{\delta'}
\end{array}
\end{equation}
We can compute the vector in the direction of any outgoing beam $\vec{q}$
using the grating equation (\ref{eq:vec_grating_comp}). From the condition
for the $y$ coordinate we can immediately compute $\delta''$: 
\begin{equation}
q_{y}=p_y=\sin{\delta'}\qquad\Rightarrow\qquad \delta''=\delta'
\end{equation}
The projection on $x$ can thus be written as:
\begin{equation}
q_x=-\sin{(\beta_m+\Delta \beta_m)}\cos{\delta'}
\approx-\left(\sin{\beta_m} + \Delta \beta_m\cos{\beta_m}\right) \cos{\delta'}\\
\end{equation}
The minus sign originates from the fact that we consider a reflection grating which turns the
direction vector; e.g. for $\alpha=\beta_m$ we must obtain $q_x=-p_x$.
Using the above and the grating equation we get:
\begin{equation}
\begin{array}{l}
\cos{\delta'}\left(\sin{\alpha}+\Delta\alpha\cos\alpha\right) +\cos{\delta'}\left(\sin{\beta_m} + \Delta \beta_m\cos{\beta_m}\right)=\frac{m \lambda}{d}\\
\end{array}
\end{equation}
For small $\delta'$ we can write:
\begin{equation}
\Delta \beta_m \approx \left(\frac{m \lambda}{d}~\frac{\delta'^2}{2}-\Delta \alpha \cos\alpha\right)/\cos\beta_m
\end{equation}

\paragraph{Rotation of the incoming beam around the $y$-axis:}
For small rotations of the incoming beam by $\Delta \alpha$ around the $y'$ axis we obtain: 
\begin{equation}
\Delta \beta_m=-\frac{\cos{\alpha}}{\cos{\beta_m}}\Delta \alpha\quad\mbox{and}\quad\delta''=0
\end{equation}

\paragraph{Rotation of the incoming beam around the $x'$-axis:}
A single rotation by $\delta'$ around the $x'$ axis yields:
\begin{equation}
\Delta \beta_m \approx \frac{m \lambda}{d}~\frac{\delta'^2}{2\cos\beta_m}\quad\mbox{and}\quad\delta''=\delta'
\end{equation}

\paragraph{Rotation of the grating around the $z$-axis:}
In order to compute the effect of a rotation of the grating around its normal by an angle $\gamma$ 
we again make use of the vectorial grating equation~(\ref{eq:vec_grating}). In this case the
unit vector along the direction of the grooves is given as $(\sin\gamma, \cos\gamma,0)$
and we obtain the following set of equations:
\begin{equation}
q_x-p_x=-\frac{m\lambda}{d}\cos\gamma\quad\mbox{and}\quad q_y-p_y=\frac{m\lambda}{d}\sin\gamma
\end{equation}
With the input beam aligned we can use the following projections $p_x=\sin\alpha$
and $p_y=0$.
This leads to:
\begin{equation}
\begin{array}{l}
q_x=-\sin(\beta_m+\Delta \beta)=\sin\alpha-\frac{m \lambda}{d}\cos\gamma\\
q_y=\sin\delta''=\frac{m \lambda}{d}\sin\gamma\\
\end{array}
\end{equation}
For small $\gamma$ we can approximate the rotation around the $x''$ axis as:
\begin{equation}\label{eq:roll1}
\delta''\approx\frac{m \lambda}{d}\gamma
\end{equation}
We can derive $\Delta\beta$ as before; for small $\gamma$ we obtain:
\begin{equation}
\Delta \beta\approx-\frac{m \lambda}{d}\frac{\gamma^2}{2\cos\beta_m}\\
\end{equation}
This shows that in first order the roll motion of the grating will couple
into a rotation of the beam around the $x''$ axis. The order of magnitude
is the same as for a tilt motion of the grating (or mirror). Thus in contrast
to a reflective element the suspension of a grating must be designed such that
the roll motion is suppressed to the same level as the other two rotational
degrees of freedom.

In addition, during installation and pre-alignment one must ensure the right
roll angle for the grating. In the case of the VIRGO detector \cite{PreAlign}, where
the pre-alignment requires to align a diffracted beam with an 
accuracy of 30\,cm to a mirror 3\,km far away, the grating would have to be
positioned correctly with:
\begin{equation}
\gamma\approx\delta''<90\,\mu{\rm rad}~\mbox{or}~6\cdot 10^{-3}\,{\rm deg} 
\end{equation} 
The precision and dynamic range of the pre-alignment control must be designed such that
any initial mis-orientation larger than this can be corrected. This can be achieved already
with technology used in current gravitational-wave detectors.

\paragraph{Frequency Change:}
It is worth noting that a change of the laser frequency will also result in a grating specific 
change in the angle of the outgoing beam. In an otherwise aligned setup this is given by:
\begin{equation}
    \sin\alpha + \sin(\beta_m +\Delta \beta_m)= \frac{mc}{d (f+\Delta f)}=\frac{mc}{df}-\frac{mc\Delta f}{2df^2}\\
\end{equation}
Hence
\begin{equation}
\Delta \beta_m=-\frac{m\lambda}{d\cos{\beta_m}}\frac{\Delta f}{f}\quad\mbox{\rm or}\quad \Delta \beta_m=-\frac{m\lambda^2}{d\,c\,\cos{\beta_m}}\Delta f
\end{equation}
with $c$ the speed of light. By using typical values for the frequency stability in
gravitational wave detectors we can show that in this case this effect can usually be neglected:
\begin{equation}
\Delta \beta_m\approx 10^{-22}\,{\rm rad}~\left(\frac{\Delta f}{1\,{\rm Hz}}\right)\left(\frac{\lambda}{1\,\mu{\rm m}}\right)
\end{equation}

\section{Optical path length}

The optical path length is neither affected by the alignment of the grating
nor by a translation of the grating along the $y$ axis.
It only shows a dependence on $\Delta x$ and $\Delta z$.

\begin{figure}[h]
 \centerline{\includegraphics[scale=0.45]{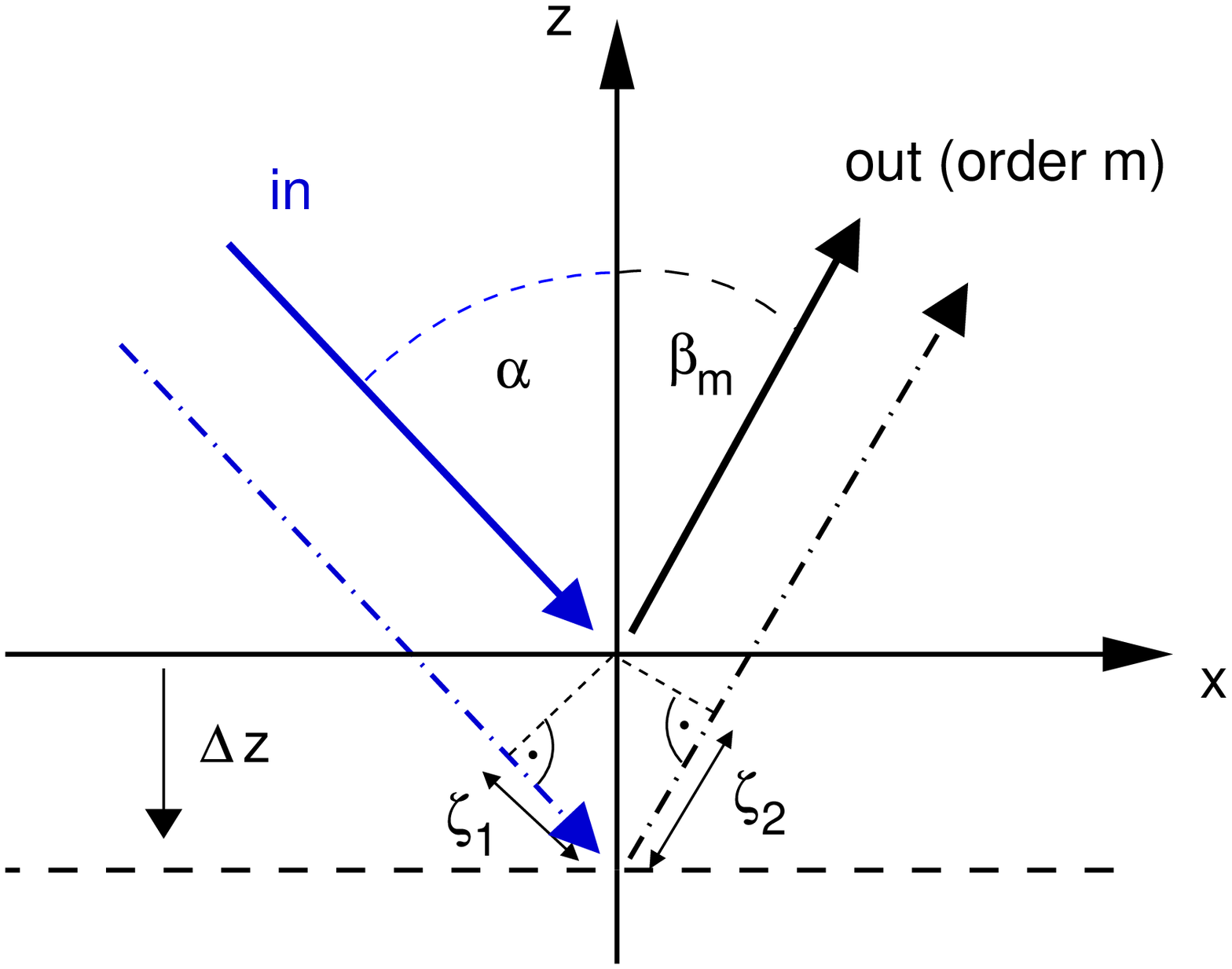}}

\vspace{10mm}
 \centerline{\includegraphics[scale=0.45, viewport=0 -60 550 320]{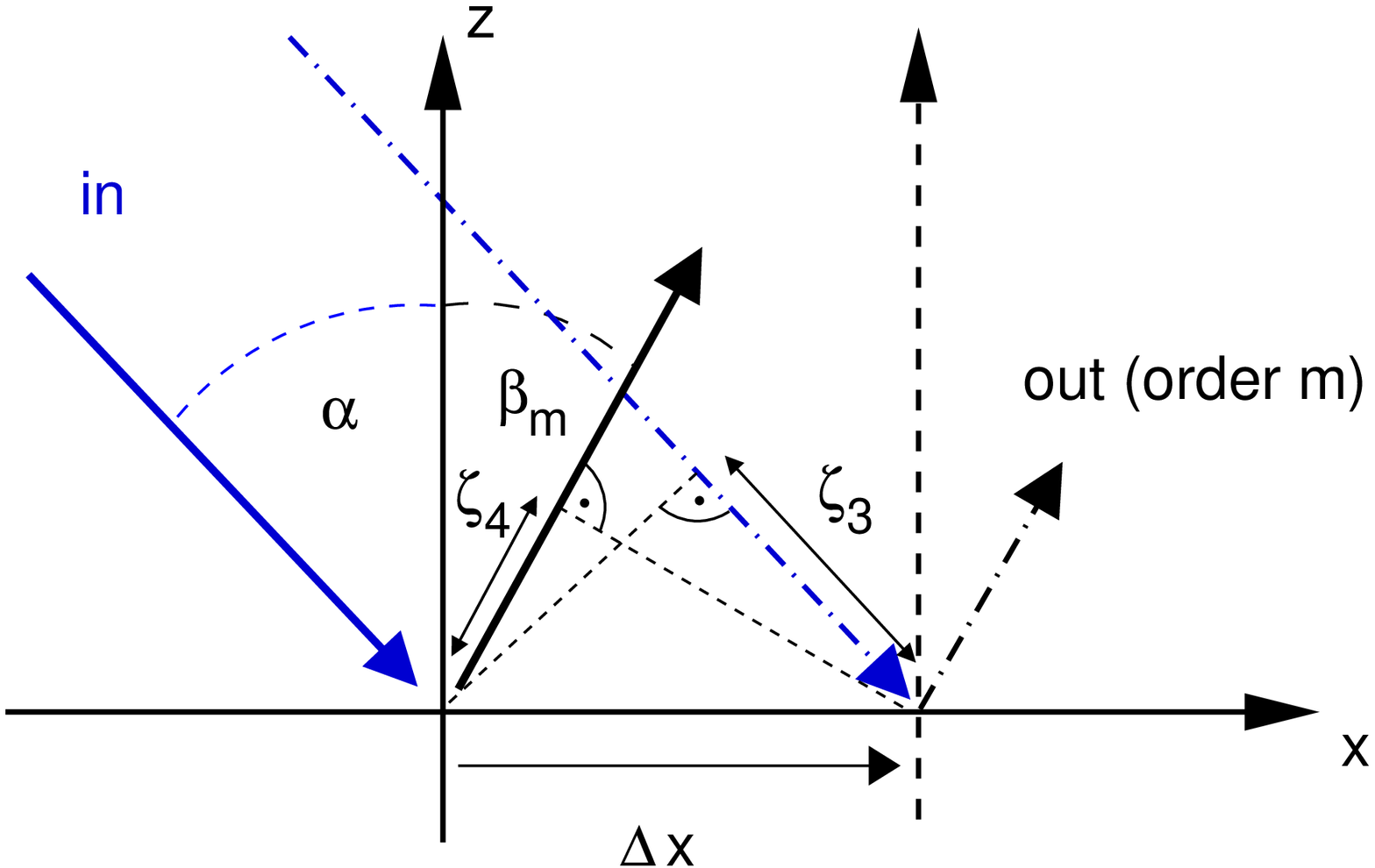}}
    \caption{Optical path length changes $\zeta$ due to translations
     of the grating. The top schematic illustrates the translation of the
     grating along $z$, the bottom schematic depicts a grating translated
     along the $x$-axis (please note that the grating itself is omitted in these
     figures; only the respective coordinates are shown). The optical path length difference with respect
     to either translation follows directly from the optical geometry
     (see text). Please note that all angles and auxiliary variables ($\Delta x$, $\zeta_3$, \dots) carry signs
     and are not defined as mere distances. For example, in the bottom schematic
     $\beta_m$ and $\zeta_4$ are negative as well as $\Delta z$ in the top graphic.}\label{fig:grphase1}
\end{figure}

The optical path length change following a translation of the
grating by $\Delta z$ can be computed from the geometry alone,
see Figure~\ref{fig:grphase1}:
\begin{equation}
\zeta_{\Delta z}=\zeta_1 + \zeta_2=-\Delta z~({\cos{\alpha}}+{\cos{\beta_m}})
\end{equation}
The minus sign reflects the definition of the phase change: The optical path
length must become larger when the grating is moved towards smaller $z$ ($\Delta z<0$).

Also the translation of the grating along the $x$-axis introduces a change in the
optical path length. This phase change is rather counter-intuitive \cite{Wise05}
but follows similarly from the geometry of the problem.
The bottom plot in Figure~\ref{fig:grphase1} shows two parallel rays diffracted by the
grating. These rays are understood to be components of the same plane wave.
By definition both rays have the same phase (modulo $\lambda$) in every reference
plane perpendicular to their direction of propagation. However, if we assume
no (or a constant) phase change at the grating surface and follow the rays
through the system to a reference plane in the outgoing field
we will obtain a phase difference $\Delta \phi$ between the two rays of
\begin{equation}\label{eq:grphase1}
\Delta \phi \frac{\lambda}{2\pi}=\zeta_3 + \zeta_4=\Delta x~({\sin{\alpha}}+{\sin{\beta_m}})
\end{equation}
(please note that $\zeta_4$ as shown in Figure~\ref{fig:grphase1} would be negative).
This has two implications for the phase of the diffracted beam:
\begin{itemize}
\item
the diffraction at the grating must advance or retard the phases for those parts of the
plane wave that hit the grating with a spatial distance $\Delta x$ to a chosen reference.

We can follow that the grating introduces a change in the optical path length as:
\begin{equation}\label{eq:grphase1b}
\zeta_{\Delta x}=-\Delta x~({\sin{\alpha}}+{\sin{\beta_m}})
\end{equation}
\item
We can for example define the center of the incoming wave such that the phase of
the center ray
can be computed by following the ray path through the system. A lateral
displacement of the incident wave will then change the phase of the outgoing
beam by the same amount $\zeta_{\Delta_x}$ with $\Delta x$ as the displacement of
the wave projected on the grating surface.
\end{itemize}
Using the grating equation we can also write (\ref{eq:grphase1}) as:
\begin{equation}\label{eq:grphase2}
\zeta_{\Delta x}=-\Delta x~\frac{m \lambda}{d}
\end{equation}
It should be clear that a translation of the grating by $\Delta x$ is
equivalent to a translation of the incoming wave. Hence we can conclude
that a translation of the grating produces also a change of the optical
path length as stated in (\ref{eq:grphase2}).
However, due to the periodic symmetry of the grating all measurable quantities must be
identical for $\Delta x=0$ and $\Delta x=n\cdot d$ with $n$ an integer. Thus,
in this case, the change in the optical path length is periodic with
Equation~\ref{eq:grphase2} being defined for a translation of less
than one grating period, i.e. $\Delta x~\lambda/d$ is periodic with a period of $d$.

It is possible to find an eigenvector such that for a translation of the grating along this
vector the change of the optical path length is zero \cite{rushford06}. By comparing
the gradient of the path length change for translations parallel to $x$ and $z$ we 
find a vector along which the change in optical path length cancels ($\zeta_{\Delta x}+\zeta_{\Delta z}=0$).

\begin{figure}[h]
\centerline{\includegraphics[scale=.5]{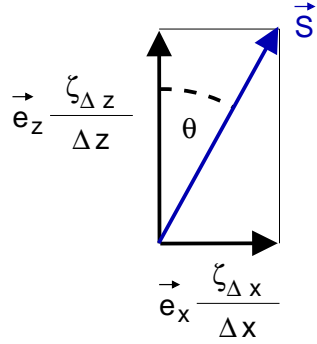}}
\caption{Eigenvector along which the translation of the grating 
does not change the optical path length (of a given refraction order):
The unit vectors $\vec{e}_x$ and $\vec{e}_z$ represent the coordinate system of the grating.
$\zeta_{\Delta x}$ and $\zeta_{\Delta z}$ refer to the optical path length change
induced by a translations by $\Delta x$ and $\Delta z$ along the respective axis (note that the
change in optical path length depends on the refraction order, see Equation~\ref{eq:grphase2}).
Thus a translation perpendicular to the vector $\vec{S}$ will yield no overall 
change in optical path length as $\zeta_{\Delta x}$ and $\zeta_{\Delta z}$
compensate exactly.}\label{fig:eigenvector}
\end{figure}
We take the ratio between the slopes of the optical path length change for $\Delta x$ and $\Delta y$:
\begin{equation}
\frac{\zeta_{\Delta x}/\Delta x}{ \zeta_{\Delta z}/\Delta z}=\frac{{\sin{\alpha}}+
{\sin{\beta_m}}}{{\cos{\alpha}}+{\cos{\beta_m}}}=\tan{\left(\frac{\alpha+\beta_m}{2}\right)}
\end{equation}
The unit vector for motion with exactly compensating changes of the optical path length must be
perpendicular to $\vec{S}$ as shown in Figure~\ref{fig:eigenvector}; the direction of $\vec{S}$ is defined by the angle
$\theta=(\alpha+\beta_m)/2$
which defines the bisection between the incoming and the diffracted beam.

Hence a translating of the grating perpendicular to the bisection of the input and 
diffracted wave vector
yields no change in the optical path length. This fact can be utilised for example
by carefully choosing a mounting and seismic isolation strategy such that the translational degree of
freedom containing the largest amount of (seismic) noise
is made perpendicular to the bisection of the incoming and one outgoing beam. However, the fact that the
axis for zero phase change can only be chosen with respect to one pair of beams and that it
cannot be perpendicular to an incoming beam makes it impossible to avoid all couplings of 
alignment noise into phase noise for higher diffraction orders, especially for a mis-alignment of 
the incoming beam.

\section{Alignment noise in exemplary optical setups}

In this section we will compute the coupling of alignment noise
into phase noise for a few simplified, exemplary optical setups. 
The optical systems are analysed in the plane perpendicular to grating
grooves, ignoring the second alignment degree of freedom of each optical component. 

We will compare the phase noise due to component or beam misalignment in standard
two-mirror cavities to that of a Fabry-Perot cavity with a grating
as the input coupler. We will further briefly discuss the 
alignment-related phase noise at beam splitters.

\subsection{Two-mirror cavity}
In this section we recall the basic geometry of a two-mirror cavity with a flat
input mirror and a spherical end mirror. The misalignment of the
optical system can be described by a misalignment of the
input mirror and/or the end mirror. Both effects result in a displacement
of the cavity eigenmode and a change of the optical path length.

\begin{figure}[h]
\centerline{\includegraphics[width=9cm]{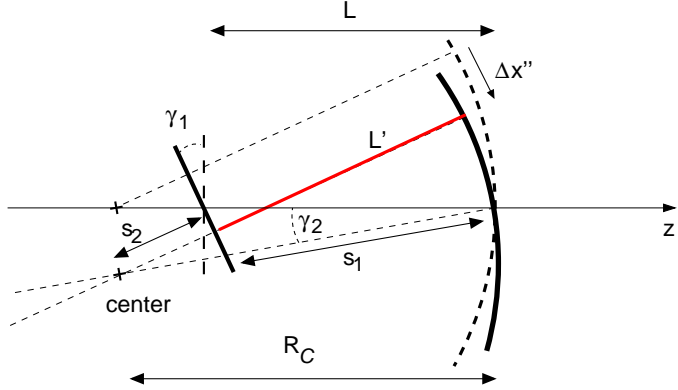}}
    \caption{The geometry of a conventional two-mirror cavity with both mirrors misaligned. The length
of the misaligned cavity $L'=R_c-s_2$ can be computed from the shown parameters as described in the text.} \label{fig:align3}
\end{figure}

Figure~\ref{fig:align3} shows the geometry for our two-mirror example cavity with both mirrors being
misaligned. In practice, also the input mirror might be spherical and the center
of rotation will probably not coincide with the optical surface. However the calculations below provide 
the order of magnitude of the misalignment effects.
By using some basic geometry we see that the new cavity length is
given by $L'=R_c-s_2$. We further find that
\begin{equation}
s_1=L\frac{\sin(\pi/2-\gamma_1)}{\sin(\pi/2-(\gamma_2-\gamma_1))}=L\frac{\cos\gamma_1}{\cos(\gamma_2-\gamma_1)}
\end{equation}
and
\begin{equation}
s_2=(R_c-s_1)\cos(\gamma_2-\gamma_1)=R_c\cos(\gamma_2-\gamma_1)-L\cos\gamma_1
\end{equation}
Thus we can write the new cavity length as
\begin{equation}
\begin{array}{ccl}
L'&=&R_c-(R_c-L\frac{\cos\gamma_1}{\cos(\gamma_2-\gamma_1)})\cos(\gamma_2 -\gamma_1)\\
&=&R_c(1-\cos(\gamma_2 -\gamma_1))+L\cos\gamma_1
\end{array}
\end{equation}
or for small angles:
\begin{equation}\label{eq:alignnoise}
L'\approx L - R_c \gamma_2 \gamma_1 + \frac{R_c}{2} \gamma_2^2 +\frac{R_c-L}{2}  \gamma_1^2 
\end{equation}
This result 
shows a quadratic dependency on the mirror misalignment. 
The change in the cavity length $\Delta L=L-L'$ yields a change in the phase of the
circulating light. In the following we will refer to the term $\Delta L$ also
as \emph{phase noise} which implies some assumptions on the frequencies
of the alignment fluctuations with respect to the cavity linewidth, see below. 

To quantify the alignment noise coupling we can compute a limit on the fluctuation of one mirror
with respect to a phase noise sensitivity.  Due to the nature of the
seismic isolation systems the alignment fluctuations of the core optical components in a 
gravitational wave detector have similar
spectral distributions: the alignment fluctuations are largest at Fourier frequencies below a 
cut-off frequency $f_c$ (often 1\,Hz) and decrease rapidly with increasing frequency. Thus 
the phase noise $\Delta L$
at a given frequency $f_n>f_c$ as a function of the quadratic
coupling of alignment fluctuation as shown in (\ref{eq:alignnoise}) is dominated by the
mix-terms between a low-frequency (quasi-static) misalignment and the fluctuations at
$f_n$. We therefore write the 
alignment fluctuations as a 
sum of a DC term and an AC term:
\begin{equation}\label{eq:alignnoise2}
\gamma=\gamma_{\rm DC} + \gamma(f_n)
\end{equation}
This yields for the phase noise at a given Fourier frequency $f_n>f_c$:
\begin{equation}\label{eq:alignnoise3}
\begin{array}{lcl}
\Delta L(f_n)&=&R_c (\gamma_{\rm 2,DC}\,\gamma_1(f_n)+\gamma_{\rm 1,DC}\,\gamma_2(f_n))\\
&& + R_c \gamma_{\rm 2,DC}\,\gamma_2(f_n) + (R_c-L)\gamma_{\rm 1,DC}\,\gamma_1(f_n)\\
\end{array}
\end{equation}
This change in cavity length is equivalent to that of a mirror displacement of $\Delta L$ if the frequency
of the alignment fluctuations are within the cavity linewidth ${\rm FWHM} > 2 f_n$.
In general the frequencies of interest are largely outside the cavity bandwidth. However, 
the low frequency limit of the measurement band can be considered to be within the
cavity linewidth (limits for third generation ground based detectors are expected to be 
between 1\,Hz and 50\,Hz). Alignment noise specification are of special interest at
this lower bound because the alignment fluctuations fall steeply with increasing frequency.
The following alignment noise limits will be computed for Fourier frequencies
$f_n$ in the low frequency band of the detector sensitivity.

Commonly the alignment noise specifications are computed such that a certain amount of quasi-static 
misalignment is assumed,
based on experience with suspension control systems, 
and then a limit for the high frequency fluctuations can
be derived with respect to a given sensitivity limit.
A typical value for the residual root mean
square misalignment integrated over a band between 0\,Hz to 10\,Hz for future gravitational-wave 
detectors can be assumed to be $\gamma_{\rm DC}=10\,$nrad \cite{Mueller05}.
We consider the mirrors to be misaligned statically by $\gamma_{\rm 1,DC}$ and $\gamma_{\rm 2,DC}$ and 
the far mirror is rotating further with an amplitude of $\gamma_2(f_n)$ at frequency $f_n$.
This yields a phase noise of:
\begin{equation}\label{eq:mixterm2}
\begin{array}{lcl}
\Delta L_{\gamma_2}(f_n)&=&R_c\,\gamma_{\rm 1,DC}\, \gamma_2(f_n)- R_c\,\gamma_{\rm 2,DC}\, \gamma_2(f_n)\\
\end{array}
\end{equation}
Equation \ref{eq:mixterm2} shows that the
maximum phase noise is reached when $\gamma_{\rm 2,DC}=-\gamma_{\rm 1,DC}$, which yields:
\begin{equation}\label{eq:mixterm3}
\begin{array}{lcl}
\Delta L_{\gamma_2}(f_n)&=&2 R_c\,\gamma_{\rm 1,DC}\, \gamma_2(f_n)
\end{array}
\end{equation}
If we take $\Delta L_{\gamma_2}(f_n)$ to be the (differential) phase noise in the arm cavities of a Michelson
interferometer and we further assume a sensitivity goal for that Michelson
of $h=10^{-23}/\sqrt{\rm Hz}$ we can use (\ref{eq:mixterm3}) to compute a limit for 
$\gamma_s(f_n)$, i.e. the alignment noise at the second mirrors.
Using exemplary parameters for the cavity length and mirror curvature from the 
VIRGO interferometer \cite{virgo2006short}, we obtain the following alignment noise limit:
\begin{equation}\label{eq:normallimit}
\fl\gamma_2(f_n) < 2\cdot 10^{-16}\,\frac{\rm rad}{\sqrt{\rm Hz}}~ \left(\frac{h}{10^{-23}/\sqrt{\rm Hz}}\right)\left(\frac{L}{3\,\mbox{km}}\right)\left(\frac{3.5\,\mbox{km}}{R_c}\right)\left(\frac{10\,\mbox{nrad}}{\gamma_{\rm 1, DC}}\right)
\end{equation} 

In the following sections we will use the same method to compute limits on alignment fluctuations
for cavities that employ gratings. In the presence of gratings we also need
to know the displacement of the optical axis.
In order to compute the lateral translation of the eigenmode with both mirrors misaligned we first compute the $x$ 
coordinate of the center of the sphere associated with the end mirror:
\begin{equation}
x_c=-R_c\sin\gamma_2
\end{equation}
With this we can compute the $x$ coordinate of the point where the eigenmode touches the front mirror:
\begin{equation}
\fl x_f=x_c+s_2\sin\gamma_1=\sin\gamma_1(R_c\cos(\gamma_2-\gamma_1)-L\cos\gamma_1)-R_c\sin\gamma_2
\end{equation}
which gives a new $x''$ coordinate for the eigenmode of
\begin{equation}
x''=\frac{x_f}{\cos\gamma_1}=R_c\sin(\gamma_2 +\gamma_1)-L\sin\gamma_1
\end{equation}
And for small angles
\begin{equation}
x''\approx R_c\gamma_2 + (R_c-L)\gamma_1
\end{equation}
We find that
the lateral displacement is linearly dependent on the misalignment angles.
This displacement does not result in a dominant phase noise contribution when
conventional mirrors are used but is shown in the next section 
to be critical when gratings are employed.

\subsection{Two-port grating as coupling 'mirror' into arm cavity}

In this section we will compare the above results to those of a similar cavity with a grating 
as the input couplers (see Figure~\ref{fig:ifos} bottom, left). 
We use the same parameters as above; i.e. the grating is flat and the end mirror is spherical.
It should be clear that the cavity with a grating experiences exactly the same 
coupling of alignment into phase noise as computed for the two-mirror cavity. However,
there is an additional coupling process through the transversal displacement
of the eigenmode on the grating.

The displacement of the optical axis for a misalignment of 
the end mirror by $\gamma_2(f_n)$ can be approximated as:
\begin{equation}
\Delta x''_{\gamma_2}(f_n)=\Delta x'\approx R_C \gamma_2(f_n)
\end{equation}
With respect to the grating coordinate system we shall write:
\begin{equation}
\Delta x=\frac{\Delta x'}{\cos(\alpha)}=\frac{ R_C}{\cos(\alpha)}~ \gamma_2
\end{equation}
As shown above, the translation of the beam on the grating will result in a variation of the optical path length
during the refraction as:
\begin{equation}
\Delta \zeta=\Delta x \frac{m \lambda}{d}=\frac{ R_C}{\cos(\alpha)}\frac{\lambda}{d}~ \gamma_2
\end{equation}
Such change in phase corresponds to an apparent fluctuation in the cavity length
of $\Delta L = 0.5 \Delta \zeta$.
From this we can compute, for example, new limits for the alignment noise of the second mirror. Assuming 
again a sensitivity goal
of $h=10^{-23}$, VIRGO-like parameters and typical values for the grating parameters we can write:
\begin{equation}\label{eq:limit2}
\fl \gamma_2 < 7\cdot 10^{-24}\,\frac{\rm rad}{\sqrt{\rm Hz}}~ \left(\frac{h}{10^{-23}/\sqrt{\rm Hz}}\right)\left(\frac{L}{3\,\mbox{km}}\right)\left(\frac{\cos(\alpha)}{\cos(30^\circ)}\right) \left(\frac{3.5\,\mbox{km}}{R_C}\right)\left(\frac{d}{\lambda}\right)
\end{equation}
which proves to be a much more stringent alignment requirement than the direct coupling mechanism of alignment into phase noise given
in (\ref{eq:normallimit}).

\subsection{Three-port grating as coupling 'mirror' into arm cavity}
In the case of a three-port grating as the cavity input mirror with the cavity mode
impinging on the grating at normal incidence the situation is a little different:
Since the cavity mode does not experience refraction at the grating but rather
a zero order reflection, a translation of the mode will not create an optical
path length change within the cavity. However, the beams leaving or entering the cavity
will experience exactly the same phase noise as described above. The main difference is
that the limits in this case are relaxed by the finesse of the arm cavity.
Hence, for a cavity with finesse $F$ and a three-port grating coupling mirror
we can compute limits for the misalignment of the far mirror to be:
\begin{equation}\label{eq:limit3}
\fl \gamma_2 < 1\cdot 10^{-21} \frac{\rm rad}{\sqrt{\rm Hz}} \left(\frac{h}{10^{-23}/\sqrt{\rm Hz}}\right)\left(\frac{L}{3\,\mbox{km}}\right)\left(\frac{F}{200}\right)\left(\frac{\cos(\alpha)}{\cos(45^\circ)}\right) \left(\frac{3.5\,\mbox{km}}{R_C}\right)\left(\frac{d}{\lambda}\right)
\end{equation}
However, the presence of the grating will result also in more stringent requirements for the input beam jitter 
\cite{Mueller05} which are beyond the
scope of this article.

\subsection{Four-port grating and beam splitters}
When a four-port grating is used for a beam splitter (as shown in Figure~\ref{fig:ifos}) the outgoing beams
are given by the interference between a zero-order and a first-order diffracted beam.  
Any translation of the grating or the incoming beams along their respective $x$-axis results in
phase noise in the first-order beams causing the interference to be 
directly affected by the translation. The effect is comparable to a translation
of a standard beam splitter along its surface normal.
If a four-port grating was used as a central beam splitter in a VIRGO-like
optical layout any translation of the beams impinging on this beam splitter 
would be caused primarily by changes in the optical axes of the arm cavities as computed above. 
We can thus derive alignment specifications for the arm cavity mirrors, and again, the specifications
are relaxed by the cavity finesse because the phase noise originates outside the arm cavities.
In the example of the VIRGO interferometer this translates into the same alignment limits for the far mirror
as given in Equation~\ref{eq:limit3}.

\section{Conclusion}
Diffraction gratings have been proposed as replacements of traditional mirrors and beam splitters
for interferometric gravitational-wave detectors. However, so far only draft optical layouts
have been published without an in-depth analysis of their noise performance. 
To our knowledge we have for the first time presented the effects of beam and grating alignment
on the outgoing beam in a form required to estimate the sensitivity and performance
of a long-baseline laser interferometer with reflective diffraction gratings as core optical
elements.
Diffraction gratings differ from traditional mirrors and beam splitters in several ways; in particular
they reduce the symmetry between the interacting beams. Comparing ideal diffraction
gratings with traditional, ideal mirrors and beam splitters shows that the reduced 
symmetry results in extra coupling of geometry changes of the grating or the incoming beam into 
alignment and phase changes of the outgoing beam. 
In particular, a displacement of the grating along the $x$-axis (perpendicular 
to the grating normal and to the grating grooves, see Figure~\ref{fig:coordinates}) introduces a periodic
change of the optical path length while a displacement of the incoming beam along
$x'$ (perpendicular to the beam axis and to the grating grooves) introduces a continuous change of the optical path length.
The optical path
length change is proportional to the order of the diffracted beam, especially
it is zero for the zeroth order.

The extra alignment changes are of the
same magnitude and quality as the normal alignment effects. The additional
coupling of a roll motion into beam misalignment probably requires a careful design of the suspension 
system of diffraction gratings. 
However, the additional coupling of beam alignment noise into phase noise at a grating results
in much more stringent alignment specification for main interferometer components if a
grating is used either as coupling mirror for arm cavities or as the main beam splitter.
By analysing a simplified example, using VIRGO-like parameters for the optical system, we could show that the 
currently proposed draft topologies for the use of diffraction gratings would result in challenging requirements
for the alignment of the optical components - the grating as well as the other main interferometer mirrors.
Even considering the ongoing development of suspension systems \cite{adligosusp-short} for the core optical elements
of future gravitational-wave detectors, the all-reflective topologies discussed so far would very likely
be limited by alignment noise. In order to benefit from advantages of diffraction gratings, these optical 
layouts of refractive interferometers must be designed carefully, and topologies with higher symmetry found,
in order to minimise  the alignment related phase noise. 
Furthermore, new suspension systems should be investigated, which could provide
a reduction of alignment noise to the required level.

\section{Acknowledgment}
We thank S. Hild, R. Schilling, S. Chelkowski and M. Mantovani for useful discussions. A.~F. would like to thank
STFC for financial support of this work. We also thank the Deutsche Forschungsgemeinschaft (DFG)
and the SFB TR7 for support. This document has been assigned the LIGO Laboratory document 
number LIGO-P070094-00-Z. 


\section*{References}
\bibliographystyle{unsrt}
\bibliography{gratingbib}

\end{document}